\documentclass[11pt]{article}
\usepackage[dvips]{graphicx}
\usepackage{amsmath,amssymb}
\usepackage{amscd}
\usepackage[all]{xy}

\newtheorem{Def}{Definition}
\newtheorem{Thm}[Def]{Theorem}
\newtheorem{Lem}[Def]{Lemma}

\newcommand{\Tot}{\Leftrightarrow}

\begin{document}
\title{Permutation Excess Entropy and Mutual Information between the Past and Future}

\author{Taichi Haruna\footnote{Corresponding author}$\ ^{\rm ,1,2}$,  Kohei Nakajima$\ ^{\rm 3}$ \\
\footnotesize{$\ ^{\rm 1}$ Department of Earth \& Planetary Sciences, Graduate School of Science, } \\
\footnotesize{Kobe University, 1-1, Rokkodaicho, Nada, Kobe, 657-8501, JAPAN} \\
\footnotesize{$\ ^{\rm 2}$ PRESTO, Japan Science and Technology Agency (JST),} \\
\footnotesize{4-1-8 Honcho Kawaguchi, Saitama 332-0012, Japan} \\
\footnotesize{$\ ^{\rm 3}$ Artificial Intelligence Laboratory, Department of Informatics, } \\
\footnotesize{University of Zurich, Andreasstrasse 15, 8050 Zurich, Switzerland} \\
\footnotesize{E-mail: tharuna@penguin.kobe-u.ac.jp (T. Haruna)} \\
\footnotesize{Tel \& Fax: +81-78-803-5739} \\
}

\date{}
\maketitle

\begin{abstract}
We address the excess entropy, which is a measure of complexity for stationary time series, 
from the ordinal point of view. We show that the permutation excess entropy is equal to 
the mutual information between two adjacent semi-infinite blocks in the space of orderings 
for finite-state stationary ergodic Markov processes. This result may shed a new light on the 
relationship between complexity and anticipation. 
\end{abstract}
{\bf Keywords:} Permutation Entropy; Excess Entropy; Mutual Information; Duality

\section{Introduction}
Recently, it was found that much of the information contained in stationary time series 
can be captured by orderings between values, not the values themselves \cite{Amigo2010}. 
The permutation entropy rate which was first introduced in \cite{Bandt2002a,Bandt2002b} 
quantifies the average uncertainty of orderings between values per time unit. 
This is in contrast to the usual entropy rate which quantifies the average uncertainty of 
values per time unit. However, surprisingly, it is known that the permutation entropy rate 
is equal to the entropy rate for finite-state stationary stochastic processes \cite{Amigo2010,Amigo2005}. 
Similar results for dynamical systems are also known \cite{Amigo2005,Amigo2007,Bandt2002b,Keller2010,Misiurewicz2003}. 

In our previous work \cite{Haruna2011}, we found 
a new proof of the equality between the permutation entropy rate and the entropy rate 
based on a duality between values and orderings, which can be seen as 
a Galois connection \cite{Davey2002} (categorical adjunction \cite{MacLane1998} for partially ordered sets, 
however, we do not refer to the Galois connection explicitly in this paper). 
By making use of the duality, we also proved that the permutation excess entropy is equal to 
the excess entropy for finite-state stationary ergodic Markov processes. 
The excess entropy has attracted interest from the complex systems community for decades \cite{Arnold1996,Bialek2001,Crutchfield1983,Crutchfield2003,Feldman2008,Grassberger1986,Li1991,Shaw1984}. 
By definition, the excess entropy is the sum of entropy over-estimates over finite length of words \cite{Crutchfield2003}. 
However, it can be expressed as the mutual information between the past and future, 
namely, the mutual information between two adjacent semi-infinite blocks of stochastic variables. 
Thus, the excess entropy can be interpreted as a measure of global correlation present in a system. 

In this paper, based on the duality between values and orderings, we show that the permutation excess entropy 
also admit a mutual information expression in the space of orderings when the process is finite-state 
stationary ergodic Markov. This result partially justifies the claim that the permutation excess entropy measures 
global correlation at the level of orderings between values present in stationary time series. 

This paper is organized as follows. 
In Section 2, we review the duality between values and orderings. 
In Section 3, we explain the permutation excess entropy. 
In Section 4, we present a proof of the claim that the permutation excess entropy has a mutual information expression 
for finite-state stationary ergodic Markov processes. 
In Section 5, we give conclusions. 

\section{Duality between Values and Orderings Explained}
Let $A_n=\{1,2,\cdots,n\}$ be a finite alphabet consisting of natural numbers from $1$ to $n$. 
We consider $A_n$ as a totally ordered set ordered by the usual `less-than-or-equal-to' relationship. 

We denote the set of all permutations of length $L \geq 1$ by $\mathcal{S}_L$. Namely, each element 
$\pi \in \mathcal{S}_L$ is a bijection on the set $\{1,2,\cdots,L\}$. 
For convenience, we denote each permutation $\pi \in \mathcal{S}_L$ by a string $\pi(1)\cdots\pi(L)$. 

For each {\it word} $s_1^L:=s_1 \cdots s_L:=(s_1,\cdots,s_L) \in A_n^L=\underbrace{A_n \times \cdots \times A_n}_{L}$ 
of length $L \geq 1$, we define its {\it permutation type} $\pi \in \mathcal{S}_L$ 
by re-ordering symbols $s_1,\cdots,s_L$ in increasing order: $s_1^L$ is of type $\pi$ if we have 
$s_{\pi(i)} \leq s_{\pi(i+1)}$ and $\pi(i) < \pi(i+1)$ when $s_{\pi(i)} = s_{\pi(i+1)}$ 
for $i =1,2,\cdots,L-1$. 
For example, $\pi(1)\pi(2)\pi(3)\pi(4)\pi(5)=3142$ for $s_1^4=2312$ because 
$s_3 s_1 s_4 s_2=1223$. 

We introduce a map $\phi: A_n^L \to \mathcal{S}_L$ that sends each word $s_1^L$ to its unique permutation type 
$\pi=\phi(s_1^L)$. This map $\phi$ classifies or coarse-grains words of length $L$ by the criterion whether 
they have the same permutation type. In general, $\phi$ is many-to-one map. For example, all of 
$111,112,122,222 \in A_2^3$ have the same permutation type $\pi \in \mathcal{S}_3$ defined by 
$\pi(1)\pi(2)\pi(3)=123$ (identity on $\{1,2,3\}$). 

Now, we list the properties of the map $\phi$ which will be used later.  

\begin{Lem}
For $s_1^L, t_1^L \in A_n^L$, $\phi(s_1^L)=\phi(t_1^L)$ if and only if 
$s_k \leq s_j \Tot t_k \leq t_j$ for all $1 \leq j \leq k \leq L$. 
\label{lem1}
\end{Lem}
{\it Proof.}
See Corollary 4 in \cite{Haruna2011}. 
\hfill $\Box$ \\

\begin{Lem}
Let $n \geq i \geq 1$. Fix $\pi \in \mathcal{S}_L$. 
Assume that there is no $s_1^L \in A_{i-1}^L$ such that $\phi(s_1^L)=\pi$, but 
there exists $s_1^L \in A_i^L$ such that $\phi(s_1^L)=\pi$ 
(When $i=1$ we define $A_{i-1}=A_0=\emptyset$). 
\begin{itemize}
\item[(i)]
There exists a unique $s_1^L \in A_i^L$ such that $\phi(s_1^L)=\pi$. 
Moreover, if $\phi(t_1^L)=\pi$ for $t_1^L \in A_n^L$, then 
there exist $c_1,\cdots,c_L$ such that 
$s_k+c_k=t_k$ for $k=1,\cdots,L$ and $0 \leq c_{\pi(1)} \leq \cdots \leq c_{\pi(L)} \leq n-i$. 
\item[(ii)]
$|\phi^{-1}(\pi)|=\binom{L+n-i}{n-i}$, where $|X|$ denotes the cardinality of a set $X$. 
\end{itemize}
\label{lem2}
\end{Lem}
{\it Proof.}
See Lemma 5 in \cite{Haruna2011}. (ii) follows from the fact that the number of sequences $a_1 \cdots a_L$ 
satisfying $0 \leq a_1 \leq a_2 \leq \cdots \leq a_L \leq n-i$ is given by a binomial coefficient 
$\binom{L+n-i}{n-i}$. 
\hfill $\Box$ \\

For example, let $\phi \in \mathcal{S}_5$ be given by $\pi(1)\pi(2)\pi(3)\pi(4)\pi(5)=24315$. 
We have $\phi(s_1^5)=\pi$ for $s_1^5=s_1 s_2 s_3 s_4 s_5=31213 \in A_3^5$. Consider 
$t_1^5=t_1 t_2 t_3 t_4 t_5=41325 \in A_5^5$ and $c_1 c_2 c_3 c_4 c_5=10112$. We have $\phi(t_1^5)=\pi$ and 
$t_2 t_4 t_3 t_1 t_5=12345=11233+01112=s_2 s_4 s_3 s_1 s_5 + c_2 c_4 c_3 c_1 c_5$. 

As a more thorough illustration of Lemma \ref{lem2}, 
let us write down how $\phi$ sends each word to its permutation type for $L=3$ and $n=1,2$. 

When $n=1$, the unique element $111 \in A_1^3$ is mapped to $123 \in \mathcal{S}_3$. 

When $n=2$, we have 
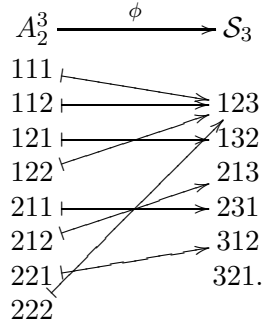
\begin{eqnarray*}
\xymatrix@R=0pt{
A_2^3 \ar[rr]^{\phi} & & \mathcal{S}_3 \\
111 \ar@{|->}[drr] & & \\
112 \ar@{|->}[rr] & & 123 \\
121 \ar@{|->}[rr] & &132 \\
122 \ar@{|->}[uurr] & & 213 \\
211 \ar@{|->}[rr] & & 231 \\
212 \ar@{|->}[uurr] & & 312 \\
221 \ar@{|->}[urr] & & 321. \\
222 \ar@{|->}[uuuuuurr] & & 
}
\end{eqnarray*}

For example, there is no $s_1^3 \in A_1^3$ suth that $\phi(s_1^3)=132 \in \mathcal{S}_3$. 
On the other hand, $\phi^{-1}(132)=\{121\}$ for $\phi: A_2^3 \to \mathcal{S}_3$. 
We have $\phi^{-1}(123)=\{111,112,122,222\}$ for $\phi: A_2^3 \to \mathcal{S}_3$. Note that 
$|\phi^{-1}(123)|=4=\binom{3+2-1}{2-1}$. 

Let us introduce a map $\mu : \mathcal{S}_L \to \mathbb{N}^L$, where $\mathbb{N}=\{1,2,\cdots\}$ is 
the set of all natural numbers, by the following procedure: 
\begin{itemize}
\item[(i)]
Given a permutation $\pi \in \mathcal{S}_L$, we decompose the sequence $\pi(1) \cdots \pi(L)$ 
into {\it maximal ascending subsequences}. A subsequence $i_j \cdots i_{j+k}$ of a sequence $i_1 \cdots i_L$ 
is called a {\it maximal ascending subsequence} if it is ascending, namely, $i_j \leq i_{j+1} \leq \cdots \leq i_{j+k}$, 
and neither $i_{j-1} i_{j} \cdots i_{j+k}$ nor $i_{j} i_{j+1} \cdots i_{j+k+1}$ is ascending. 
\item[(ii)]
If $\pi(1) \cdots \pi(i_1), \ \pi(i_1+1) \cdots \pi(i_2), \cdots, \pi(i_{k-1}+1) \cdots \pi(L)$ 
is a decomposition of $\pi(1)\cdots\pi(L)$ into maximal ascending subsequences, then we define 
a word $s_1^L \in \mathbb{N}^L$ by 
\begin{eqnarray*}
s_{\pi(1)}=\cdots=s_{\pi(i_1)}=1, s_{\pi(i_1+1)}=\cdots=s_{\pi(i_2)}=2, \cdots, s_{\pi(i_{k-1})+1}=\cdots=s_{\pi(L)}=k. 
\end{eqnarray*}
We define $\mu(\pi)=s_1^L$. 
\end{itemize}

By construction, we have $\phi \circ \mu(\pi)=\pi$ when $\mu(\pi) \in A_n^L$ 
for all $\pi \in \mathcal{S}_L$. 

For example, a decomposition of $15423 \in \mathcal{S}_5$ into maximal ascending subsequences is 
$15,4,23$. We obtain $\mu(\pi)=s_1 s_2 s_3 s_4 s_5=13321$ by putting $s_1 s_5 s_4 s_2 s_3=11233$. 

The map $\mu$ can be seen as a dual to the map $\phi$ in the following sense: 

\begin{Thm}
Let us put 
\begin{eqnarray}
X &=& \{s_1^L \in A_n^L | \phi^{-1}(\pi)=\{s_1^L\} \text{ for some } \pi \in \mathcal{S}_L \}, \label{eq1}\\
Y &=& \{\pi \in \mathcal{S}_L | |\phi^{-1}(\pi)|=1 \}. \label{eq2}
\end{eqnarray}
Then, $\phi$ restricted on $X$ is a map into $Y$, $\mu$ restricted on $Y$ is a map into $X$, 
and they form a pair of mutually inverse maps. Furthermore, we have 
\begin{eqnarray}
X=\{s_1^L \in A_n^L | 1 \leq \forall i \leq n-1 \ 1 \leq \exists j < k \leq L \text{ s. t. } s_j=i+1,s_k=i \}
\label{eq3}
\end{eqnarray}
\label{thm3}
\end{Thm}
{\it Proof.}
See Theorem 9 in \cite{Haruna2011}. 
\hfill $\Box$ \\

For the map $\phi: A_2^3 \to \mathcal{S}_3$, the duality 
\begin{eqnarray}
\xymatrix{
X \ar@<1ex>[r]^{\phi} & Y \ar@<1ex>[l]^{\mu}
}
\label{eq4}
\end{eqnarray}
is given by 
\begin{eqnarray*}
\xymatrix@R=0pt{
121 \ar@{<~>}[rr] & &132 \\
211 \ar@{<~>}[drr] & & 213 \\
212 \ar@{<~>}[urr] & & 231 \\
221 \ar@{<~>}[rr] & & 312. 
}
\end{eqnarray*}

\section{Permutation Excess Entropy}
Let ${\bf S}=\{S_1,S_2,\cdots\}$ be a finite-state stationary stochastic process, 
where each stochastic variable $S_i$ takes its value in $A_n$. 
By stationarity, we mean 
\begin{eqnarray*}
{\rm Pr}\{S_1=s_1,\cdots,S_L=s_L\}={\rm Pr}\{S_{k+1}=s_1,\cdots,S_{k+L}=s_L\}
\end{eqnarray*}
for any $k, L \geq 1$ and $s_1,\cdots,s_L \in A_n$. Hence, we can define 
the probability of occurrence of each word $s_1^L \in A_n^L$ by 
$p(s_1^L):=p(s_1 \cdots s_L):={\rm Pr}\{S_1=s_1,\cdots,S_L=s_L\}$. 

The {\it entropy rate} $h({\bf S})$ of a finite-state stationary stochastic process 
${\bf S}=\{S_1,S_2,\cdots\}$, which quantifies the average uncertainty of values per time unit, 
is defined by 
\begin{eqnarray}
h({\bf S})=\lim_{L \to \infty} \frac{1}{L} H(S_1^L), 
\label{eq5}
\end{eqnarray}
where $H(S_1^L)=H(S_1,\cdots,S_L)=-\sum_{s_1^L \in A_n^L} p(s_1^L) \log_2 p(s_1^L)$. 
The limit exists for any finite-state stationary stochastic process \cite{Cover1991}. 

The permutation entropy rate quantifies the average uncertainty of orderings between values per time unit. 
It is defined by 
\begin{eqnarray}
h^*({\bf S})=\lim_{L \to \infty} \frac{1}{L} H^*(S_1^L) 
\label{eq6}
\end{eqnarray}
if the limit exists, 
where $H^*(S_1^L)=H^*(S_1,\cdots,S_L)=-\sum_{\pi \in \mathcal{S}_L} p(\pi) \log_2 p(\pi)$ and 
$p(\pi)$ is the probability that $\pi$ is realized in ${\bf S}$, namely, $p(\pi)=\sum_{s_1^L \in \phi^{-1}(\pi)} p(s_1^L)$ for $\pi \in \mathcal{S}_L$. 

\begin{Thm}
For any finite-state stationary stochastic process ${\bf S}$, the permutation entropy rate $h^*({\bf S})$ exists and 
\begin{eqnarray}
h^*({\bf S})=h({\bf S}). 
\label{eq7}
\end{eqnarray}
\label{thm4}
\end{Thm}
{\it Proof.}
The proof appealing to ergodic theory is found in \cite{Amigo2010,Amigo2005}. 
For an alternative proof based on the duality between values and orderings, see \cite{Haruna2011}. 
\hfill $\Box$ \\

The entropy rate can be seen as a measure of randomness of a finite-state stationary stochastic process. 
Meanwhile the excess entropy can be interpreted as a measure of complexity \cite{Feldman2008}. 
More precisely, it measures global correlation present in a system. The excess entropy ${\bf E}({\bf S})$ 
of a finite-state stationary stochastic process ${\bf S}$ is defined by \cite{Crutchfield2003} 
\begin{eqnarray}
{\bf E}({\bf S})=\lim_{L \to \infty} \left( H(S_1^L) - h({\bf S})L \right) 
\label{eq8}
\end{eqnarray}
if the limit exists. If ${\bf E}({\bf S})$ exists, then we have \cite{Crutchfield2003} 
\begin{eqnarray}
{\bf E}({\bf S})=\sum_{L=1}^{\infty} \left( H(S_L|S_1^{L-1}) - h({\bf S}) \right)=\lim_{L \to \infty} I(S_1^L ; S_{L+1}^{2L}), 
\label{eq9}
\end{eqnarray}
where $H(Y|X)$ is the conditional entropy of $Y$ given $X$ and $I(X;Y)$ is the mutual information between $X$ and $Y$ 
for stochastic variables $X$ and $Y$. 

The permutation excess entropy was introduced in \cite{Haruna2011} by imitating the definition of 
the excess entropy. The permutation excess entropy ${\bf E}^*({\bf S})$ of a finite-state stationary 
stochastic process ${\bf S}$ is defined by 
\begin{eqnarray}
{\bf E}^*({\bf S})=\lim_{L \to \infty} \left( H^*(S_1^L) - h^*({\bf S})L \right), 
\label{eq10}
\end{eqnarray}
if the limit exists. However, it is unclear what form of correlation the permutation excess entropy quantifies 
from this expression. In the following discussion, we partially resolve this problem. 
We will show that the equality 
\begin{eqnarray}
{\bf E}^*({\bf S})=\lim_{L \to \infty} I(\phi(S_1^L) ; \phi(S_{L+1}^{2L})) 
\label{eq11}
\end{eqnarray}
holds for any {\it finite-state stationary ergodic Markov process} ${\bf S}$. Recall that the entropy rate 
and the excess entropy of a finite-state stationary Markov process ${\bf S}$ are given by 
$h({\bf S})=-\sum_{i,j=1}^n p_i p_{ij} \log_2 p_{ij}$ and 
${\bf E}({\bf S})=-\sum_{i=1}^n p_i \log_2 p_i + \sum_{i,j=1}^n p_i p_{ij} \log_2 p_{ij}$, respectively, 
where $P=(p_{ij})$ is a {\it transition matrix} and ${\bf p}=(p_1,\cdots,p_n)$ is a {\it stationary distribution}. 
$P$ and ${\bf p}$ satisfy $p_{ij} \geq 0$ for all $1 \leq i,j \leq n$, $\sum_{j=1}^n p_{ij}=1$ for 
all $1 \leq i \leq n$, $p_i \geq 0$ for all $1 \leq i \leq n$, $\sum_{i=1}^n p_i=1$ 
and $\sum_{i=1}^n p_i p_{ij}=p_j$ for all $1 \leq j \leq n$. 
The probability of occurrence of each word $s_1^L \in A_n^L$ is given by 
$p(s_1^L)=p_{s_1}p_{s_1 s_2} \cdots p_{s_{L-1} s_L}$. A finite-state stationary Markov process ${\bf S}$ 
is {\it ergodic} if and only if its transition matrix $P$ is {\it irreducible} \cite{Walters1982}: 
a matrix $P$ is {\it irreducible} if for all $1 \leq i,j \leq n$ there exists $l>0$ 
such that $p_{ij}^{(l)}>0$, where $p_{ij}^{(l)}$ is the $(i,j)$-th element of $P^l$. 
For an irreducible non-negative matrix, stationary distribution 
${\bf p}=(p_1,\cdots,p_n)$ exists uniquely and satisfies $p_i>0$ for all $1 \leq i \leq n$. 

In our previous work \cite{Haruna2011}, we showed that the equality 
\begin{eqnarray}
{\bf E}^*({\bf S})={\bf E}({\bf S}) 
\label{eq12}
\end{eqnarray}
holds for any finite-state stationary ergodic Markov process. The key point of the proof is 
that the probability 
\begin{eqnarray}
q_L=\sum_{\begin{subarray}{c} \pi \in \mathcal{S}_L, \\ |\phi^{-1}(\pi)|>1 \end{subarray} } p(\pi) 
=\sum_{\pi \not\in Y } p(\pi)
\label{eq13}
\end{eqnarray}
diminishes exponentially fast as $L \to \infty$ for any finite-state stationary ergodic Markov process, 
where the set $Y$ is given by (\ref{eq2}) in Theorem \ref{thm3}. 
For the proof of the equality (\ref{eq11}), we also appeal to this fact. Hence, we shortly review 
the reason why this fact follows. 

Let $L$ be a positive integer. We introduce the following probability $\beta_s$ for each symbol $s \in A_n$: 
\begin{eqnarray}
\beta_s = {\rm Pr}\{ s_1^N | s_j \neq s \text{ for any } 1 \leq j \leq N \}, 
\label{eq14}
\end{eqnarray}
where $N=\lfloor L/2 \rfloor$ and $\lfloor x \rfloor$ is the largest integer not greater than $x$. 

\begin{Lem}[Lemma 12 in \cite{Haruna2011}]
Let ${\bf S}$ be a finite-state stationary stochastic process and 
$\epsilon$ be a positive real number. 
If $\beta_s \leq \epsilon$ for any $s \in A_n$, then $q_L \leq 2n \epsilon$. 
\label{lem5}
\end{Lem}
{\it Proof.}
We shall prove $\sum_{\pi \in Y } p(\pi) \geq 1-2n \epsilon$, where the set $Y$ is given by (\ref{eq2}) in 
Theorem \ref{thm3}. 
Let us consider a word $s_1^L \in A_n^L$ satisfying the following two conditions: 
\begin{itemize}
\item[(i)]
Each symbol $s \in A_n$ appears in $s_1^N$ at least once. 
\item[(ii)]
Each symbol $s \in A_n$ appears in $s_{N+1}^L$ at least once. 
\end{itemize}
By the assumption of the lemma, we have 
\begin{eqnarray*}
{\rm Pr}\{ s_1^N | \text{(i) holds} \} \geq 1-n\epsilon, 
\end{eqnarray*}
because 
\begin{eqnarray*}
{\rm Pr}\{ s_1^N | \text{(i) holds} \} 
+ \sum_{s=1}^n {\rm Pr}\{ s_1^N | s_j \neq s \text{ for any } 1 \leq j \leq N \} \geq 1. 
\end{eqnarray*}
Similarly, 
\begin{eqnarray*}
{\rm Pr}\{ s_{N + 1}^L | \text{(ii) holds} \} \geq 1-n\epsilon 
\end{eqnarray*}
holds because of the stationarity. Hence, we obtain 
\begin{eqnarray*}
{\rm Pr}\{ s_1^L | \text{both (i) and (ii) hold} \} \geq 1-2n\epsilon. 
\end{eqnarray*}
Since a word $s_1^L \in A_n^L$ satisfying both (i) and (ii) is a member of the set $X$ given by (\ref{eq1}) 
in Theorem \ref{thm3}, we obtain 
\begin{eqnarray*}
\sum_{\pi \in Y } p(\pi) 
=\sum_{s_1^L \in X} p(s_1^L)
\geq {\rm Pr}\{ s_1^L | \text{both (i) and (ii) hold} \} \geq 1-2n\epsilon. 
\end{eqnarray*}

\hfill $\Box$ \\

Let ${\bf S}$ be a finite-state stationary ergodic Markov process whose transition matrix is 
$P$ and stationary distribution is ${\bf p}$. We can write 
$\beta_s$ in the following form by using Markov property: 
\begin{eqnarray}
\beta_s
= \sum_{\begin{subarray}{c} s_j \neq s, \\ 1 \leq j \leq N \end{subarray} } p(s_1 \cdots p_N) 
= \sum_{\begin{subarray}{c} s_j \neq s, \\ 1 \leq j \leq N \end{subarray}} p_{s_1} p_{s_1 s_2} \cdots p_{s_{N-1} s_N}
= \langle \left(P_s \right)^{N-1} {\bf u}_s, {\bf p} \rangle, 
\label{eq15}
\end{eqnarray}
where a matrix $P_s$ is defined by 
\begin{eqnarray*}
\left( P_s \right)_{ij}=
\begin{cases}
0 & \text{ if } i=s \\
p_{ij} & \text{otherwise,}
\end{cases}
\end{eqnarray*}
a vector ${\bf u}_s=(u_1,\cdots,u_n)$ is defined by $u_i=0$ if $i=s$ otherwise $u_i=1$ and 
$\langle \cdots,\cdots \rangle$ is the usual inner product in the $n$-dimensional Euclidean space. 

We can prove that the non-negative largest eigenvalue $\lambda$ of $P_s$ is strictly less than 1 and 
absolute value of any other eigenvalue of $P_s$ is not greater than $\lambda$ by using Perron-Frobenius Theorem 
for non-negative matrices and the irreducibility of $P$ (Lemma 13 in \cite{Haruna2011}). 
Hence, by decomposing $P_s$ into a sum of a diagonalizable matrix and a nilpotent matrix, 
we obtain the following lemma: 
\begin{Lem}
Let ${\bf S}$ be a finite-state stationary ergodic Markov process. 
There exists $0 \leq \alpha <1$, $C>0$ and a positive integer $k$ 
such that $\beta_s \leq C \alpha^L L^k$ for any $s \in A_n$ and sufficiently large $L$. 
\label{lem6}
\end{Lem}

\section{Mutual Information Expression of Permutation Excess Entropy}
In this section, we give a proof of the equality (\ref{eq11}) for finite-state stationary ergodic Markov processes. 
We make use of the notions of {\it rank sequences} and {\it rank variables} which are introduced in \cite{Amigo2005}. 

{\it Rank sequences} of length $L$ are words $r_1^L \in \mathbb{N}^L$ satisfying 
$1 \leq r_i \leq i$ for $i=1,\cdots,L$. We denote the set of all rank sequences of length $L$ 
by $\mathcal{R}_L$. Clearly, $|\mathcal{R}_L|=L!=|\mathcal{S}_L|$. 

We can transform each word $s_1^L \in A_n^L$ into a rank sequence $r_1^L \in \mathcal{R}_L$ by 
defining 
\begin{eqnarray}
r_i=\sum_{j=1}^i \delta(s_j \leq s_i), \ i=1,\cdots,L, 
\label{eq16}
\end{eqnarray}
where $\delta(X)=1$ if the proposition $X$ is true, otherwise $\delta(X)=0$. Namely, 
$r_i$ is the number of indices $j \ (1 \leq j \leq i)$ such that $s_j \leq s_i$. 
Thus, we obtain a map $\varphi:A_n^L \to \mathcal{R}_L$ such that $\varphi(s_1^L)=r_1^L$. 

We can show that the map $\varphi:A_n^L \to \mathcal{R}_L$ is compatible with the map 
$\phi:A_n^L \to \mathcal{S}_L$. Namely, there exists a bijection $\iota:\mathcal{R}_L \to \mathcal{S}_L$ 
satisfying $\iota \circ \varphi=\phi$ \cite{Haruna2011}. 

Given a stationary stochastic process ${\bf S}=\{S_1,S_2,\cdots\}$, its associated {\it rank variables} 
are defined by $R_i= \sum_{j=1}^n \delta \left( S_j \leq S_i \right)$ for $i=1,2,\cdots$. 
Note that rank variables $R_i$ $(i=1,2,\cdots)$ are not stationary stochastic variables in general. 
By the compatibility between $\phi$ and $\varphi$, we have 
\begin{eqnarray}
H(R_1^L)=H^*(S_1^L)=H(\phi(S_1^L))
\label{eq17}
\end{eqnarray}
for $L \geq 1$. 

Now, let ${\bf S}$ be a finite-state stationary ergodic Markov process. 
By (\ref{eq12}), we know that the permutation excess entropy ${\bf E}^*({\bf S})$ exists. 
By (\ref{eq17}) and chain rule, we have 
\begin{eqnarray}
{\bf E}^*({\bf S})=\lim_{L \to \infty} \left( H^*(S_1^L) - h^*({\bf S})L \right)
&=&\lim_{L \to \infty} \left( H(R_1^L) - h^*({\bf S})L \right) \\ \nonumber
&=&\sum_{L=1}^{\infty} \left( H(R_L|R_1^{L-1}) - h^*({\bf S}) \right). 
\label{eq18}
\end{eqnarray}
Since the infinite sum in (\ref{eq18}) converges, we obtain 
\begin{eqnarray}
\left| H(R_{L+1}^{2L}|R_1^L) - h^*({\bf S})L \right| 
= \left| \sum_{i=1}^{L} \left( H(R_{L+i}|R_1^{L+i-1}) - h^*({\bf S}) \right) \right| \underset{L \to \infty}{\to} 0. 
\label{eq19}
\end{eqnarray}

By the definition of mutual information, we have 
$I(\phi(S_1^L);\phi(S_{L+1}^{2L}))=H(\phi(S_{L+1}^{2L}))-H(\phi(S_{L+1}^{2L})|\phi(S_{1}^{L}))$. 
By stationarity of ${\bf S}$, $H(\phi(S_{L+1}^{2L}))=H(\phi(S_1^L))=H^*(S_1^L)$. Hence, 
it is sufficient to show that 
\begin{eqnarray}
\left| H(\phi(S_{L+1}^{2L})|\phi(S_1^L)) - h^*({\bf S})L \right| \underset{L \to \infty}{\to} 0 
\label{eq20}
\end{eqnarray}
to prove the equality (\ref{eq11}). However, by (\ref{eq19}), this reduces to showing that 
\begin{eqnarray}
\left| H(\phi(S_{L+1}^{2L})|\phi(S_1^L)) - H(R_{L+1}^{2L}|R_1^L) \right| \underset{L \to \infty}{\to} 0, 
\label{eq21}
\end{eqnarray}
which is equivalent to showing that 
\begin{eqnarray}
\left| H(\phi(S_1^L), \phi(S_{L+1}^{2L})) - H(\phi(S_1^{2L})) \right| \underset{L \to \infty}{\to} 0 
\label{eq22}
\end{eqnarray}
by (\ref{eq17}). 

\begin{Lem}
For $s_1^{2L},t_1^{2L} \in A_n^{2L}$, if $\phi(s_1^{2L})=\phi(t_1^{2L})$, then 
$\phi(s_1^{L})=\phi(t_1^{L})$ and $\phi(s_{L+1}^{2L})=\phi(t_{L+1}^{2L})$. 
Namely, the partition of $A_n^{2L}$ by the map $\phi: A_n^{2L} \to \mathcal{S}_{2L}$ is 
a refinement of the partition of $A_n^L \times A_n^L = A_n^{2L}$ by the map 
$\phi \times \phi: A_n^L \times A_n^L \to \mathcal{S}_L \times \mathcal{S}_L$. 
\label{lem7}
\end{Lem}
{\it Proof.}
The claim follows immediately from Lemma \ref{lem1}. 

\hfill $\Box$ \\

\begin{Lem}
\begin{eqnarray}
0 &\leq& H(\phi(S_1^{2L})) - H(\phi(S_1^L), \phi(S_{L+1}^{2L})) \\ \nonumber
&\leq& \left( \sum_{\begin{subarray}{c} \pi',\pi'' \in \mathcal{S}_L, \\ |\phi^{-1}(\pi')|>1 \text{ or } |\phi^{-1}(\pi'')|>1 \end{subarray}} p(\pi',\pi'') \right) 2n \log_2 (L+n)
\label{eq23}
\end{eqnarray}
holds for any finite-state stationary stochastic process ${\bf S}$, where 
\begin{eqnarray*}
p(\pi',\pi'')=\sum_{\begin{subarray}{c}s_1^L \in \phi^{-1}(\pi'),\\ s_{L+1}^{2L} \in \phi^{-1}(\pi'') \end{subarray}} p(s_1^{2L}) 
\end{eqnarray*}
for $\pi',\pi'' \in \mathcal{S}_L$. 
\label{lem8}
\end{Lem}
{\it Proof.}
By Lemma \ref{lem7}, we can write 
\begin{eqnarray*}
&& H(\phi(S_1^{2L})) - H(\phi(S_1^L), \phi(S_{L+1}^{2L})) \\
&=& -\sum_{\pi \in \mathcal{S}_{2L}} p(\pi) \log_2 p(\pi) + \sum_{\pi',\pi'' \in \mathcal{S}_{L}} p(\pi',\pi'') \log_2 p(\pi',\pi'') \\
&=& \sum_{\pi',\pi'' \in \mathcal{S}_{L}} \left( -\sum_{ \begin{subarray}{c} \phi^{-1}(\pi) \subseteq \\ (\phi \times \phi)^{-1}(\pi',\pi'') \end{subarray}} p(\pi) \log_2 p(\pi) + p(\pi',\pi'') \log_2 p(\pi',\pi'') \right) \\
&=& \sum_{\pi',\pi'' \in \mathcal{S}_{L}} \left( -\sum_{ \begin{subarray}{c} \phi^{-1}(\pi) \subseteq \\ (\phi \times \phi)^{-1}(\pi',\pi'') \end{subarray}} p(\pi) \log_2 p(\pi) + \sum_{ \begin{subarray}{c} \phi^{-1}(\pi) \subseteq \\ (\phi \times \phi)^{-1}(\pi',\pi'') \end{subarray}} p(\pi) \log_2 p(\pi',\pi'') \right) \\
&=& \sum_{\begin{subarray}{c} \pi',\pi'' \in \mathcal{S}_{L}, \\ p(\pi',\pi'')>0 \end{subarray}} p(\pi',\pi'') \left( -\sum_{ \begin{subarray}{c} \phi^{-1}(\pi) \subseteq \\ (\phi \times \phi)^{-1}(\pi',\pi'') \end{subarray}} \frac{p(\pi)}{p(\pi',\pi'')} \log_2 \frac{p(\pi)}{p(\pi',\pi'')} \right). 
\end{eqnarray*}
By Lemma \ref{lem2} (ii), we have 
\begin{eqnarray*}
0 \leq -\sum_{ \begin{subarray}{c} \phi^{-1}(\pi) \subseteq \\ (\phi \times \phi)^{-1}(\pi',\pi'') \end{subarray}} \frac{p(\pi)}{p(\pi',\pi'')} \log_2 \frac{p(\pi)}{p(\pi',\pi'')} \leq 2n \log_2 (L+n). 
\end{eqnarray*}

If $|\phi^{-1}(\pi')|=1$ and $|\phi^{-1}(\pi'')|=1$ hold for $(\pi',\pi'') \in \mathcal{S}_L \times \mathcal{S}_L$, then 
$| (\phi \times \phi)^{-1}(\pi',\pi'') |=1$. In this case, if $p(\pi',\pi'')>0$, then we have 

\begin{eqnarray*}
-\sum_{ \begin{subarray}{c} \phi^{-1}(\pi) \subseteq \\ (\phi \times \phi)^{-1}(\pi',\pi'') \end{subarray}} \frac{p(\pi)}{p(\pi',\pi'')} \log_2 \frac{p(\pi)}{p(\pi',\pi'')} =0. 
\end{eqnarray*}

\hfill $\Box$ \\

\begin{Lem}
(\ref{eq22}) holds for any finite-state stationary ergodic Markov process ${\bf S}$. 
\label{lem9}
\end{Lem}
{\it Proof.}
We have 
\begin{eqnarray*}
\sum_{\begin{subarray}{c} \pi',\pi'' \in \mathcal{S}_L, \\ |\phi^{-1}(\pi')|>1 \text{ or } |\phi^{-1}(\pi'')|>1 \end{subarray}} p(\pi',\pi'')
&\leq& \sum_{\begin{subarray}{c} |\phi^{-1}(\pi')|>1, \\ \pi'' \in \mathcal{S}_L \end{subarray}} p(\pi',\pi'')
+ \sum_{\begin{subarray}{c} |\phi^{-1}(\pi'')|>1, \\ \pi' \in \mathcal{S}_L \end{subarray}} p(\pi',\pi'') \\
&=& 2 \sum_{|\phi^{-1}(\pi')|>1} p(\pi')=2 q_L. 
\end{eqnarray*}
By Lemma \ref{lem5} and Lemma \ref{lem6}, there exist $0 \leq \alpha <1$, $C>0$ and $k>0$ such that 
$q_L \leq C \alpha^L L^k$ for sufficiently large $L$ if ${\bf S}$ is a finite-state stationary ergodic 
Markov process. The claim follows from Lemma \ref{lem8}. 

\hfill $\Box$ \\

Thus, we get our main theorem in this paper:

\begin{Thm}
The equality (\ref{eq11}) 
\begin{eqnarray*}
{\bf E}^*({\bf S})=\lim_{L \to \infty} I(\phi(S_1^L) ; \phi(S_{L+1}^{2L})) 
\end{eqnarray*}
holds for any finite-state stationary ergodic Markov process ${\bf S}$. 
\label{thm10}
\end{Thm}

\section{Conclusions}
In this paper, we showed that the permutation excess entropy is equal to the mutual information between 
the past and future in the space of orderings for finite-state stationary ergodic Markov processes. 
We hope that our result gives rise to a new insight into the relationship between complexity and anticipation. 

\subsection*{Acknowledgments}
T. Haruna was supported by JST PRESTO program.


\thebibliography{99}

\bibitem{Amigo2010}
J. M. Amig\'o, 
Permutation Complexity in Dynamical Systems. 
Springer-Verlag Berlin Heidelberg, 2010.

\bibitem{Amigo2005}
J. M. Amig\'o, M. B. Kennel, L. Kocarev, 
The permutation entropy rate equals the metric entropy rate for ergodic information sources and ergodic dynamical systems. 
Physica D 210, 77-95, 2005.

\bibitem{Amigo2007}
J. M. Amig\'o, M. B. Kennel, 
Topological permutation entropy. 
Physica D 231, 137-142, 2007.

\bibitem{Arnold1996}
D. V. Arnold, 
Information-theoretic analysis of phase transitions. 
Complex Systems 10, 143-155, 1996.

\bibitem{Bandt2002a}
C. Bandt, B. Pompe, 
Permutation entropy: a natural complexity measure for time series. 
Physical Review Letters 88, 174102, 2002.

\bibitem{Bandt2002b}
C. Bandt, G. Keller, B. Pompe, 
Entropy of interval maps via permutations. 
Nonlinearity 15, 1595-1602, 2002.

\bibitem{Bialek2001}
W. Bialek, I. Nemenman, N. Tishby, 
Predictability, complexity, and learning. 
Neural Computation 13, 2409-2463, 2001.

\bibitem{Cover1991}
T. M. Cover, J. A. Thomas, 
Elements of Information Theory. 
John Wiley \& Sons, Inc, 1991.

\bibitem{Crutchfield1983}
J. P. Crutchfield, N. H. Packard, 
Symbolic dynamics of noisy chaos. 
Physica D 7, 201-223, 1983.

\bibitem{Crutchfield2003}
J. P. Crutchfield, D. P. Feldman, 
Regularities unseen, randomness observed: Levels of entropy convergence. 
Chaos 15, 25-54, 2003.

\bibitem{Davey2002}
B. A. Davey, H. A. Priestley, 
Introduction to Lattices and Order, second edition. 
Cambridge Univ. Press, Cambridge, 2002. 

\bibitem{Feldman2008}
D. P. Feldman, C. S. McTague, J. P. Crutchfield, 
The organization of intrinsic computation: complexity-entropy diagrams and the diversity of natural information processing. 
Chaos 18, 043106, 2008. 

\bibitem{Grassberger1986}
P. Grassberger, 
Toward a quantitative theory of self-generated complexity. 
International Journal of Theoretical Physics 25, 907-938, 1986. 

\bibitem{Haruna2011}
T. Haruna, K. Nakajima, 
Permutation Complexity via Duality between Values and Orderings. 
Physica D 240, 1370-1377, 2011. 

\bibitem{Keller2010}
K. Keller, M. Sinn, 
Kolmogorov-Sinai entropy from the ordinal viewpoint. 
Physica D 239, 997-1000, 2010. 

\bibitem{Li1991}
W. Li, 
On the relationship between complexity and entropy for Markov chains and regular languages. 
Complex Systems 5, 381-399, 1991.

\bibitem{MacLane1998} 
S. MacLane, 
Categories for the Working Mathematician, second edition. 
Springer-Verlag, New York, 1998. 

\bibitem{Misiurewicz2003}
M. Misiurewicz, 
Permutations and topological entropy for interval maps. 
Nonlinearity 16, 971-976, 2003. 

\bibitem{Shaw1984}
R. Shaw, 
The Dripping Faucet as a Model Chaotic System. 
Aerial Press, Santa Cruz, California, 1984. 

\bibitem{Walters1982}
P. Walters, 
An Introduction to Ergodic Theory. 
Springer-Verlag New York, Inc, 1982.
\end{document}